# Thermal and Acoustic Measurements of Hybrid Bonding Stacks

Zirou Chen[1,#], Xinyan Xu[1,#], Zifeng Huang[1], Xufei Guo[1], Zhe Cheng[1,*]

[1] School of Integrated Circuits, Peking University, Beijing 100871, China

[#] These authors contributed equally to this work.

[*]Author to whom correspondence should be addressed: zhe.cheng@pku.edu.cn

## Abstract

Thermal and structural characterizations of hybrid bonding structures are essential for understanding heat transport in three-dimensional integrated circuits (3D-ICs). In this work, we investigate the effective thermal conductivity of the Cu/dielectric HB layer using time-domain thermoreflectance (TDTR) combined with picosecond ultrasonic technique. Shallow-angle polishing enabled depth-resolved thermal measurements by exposing hybrid bonding layers with different remaining thicknesses, while acoustic echoes were used to identify buried interfaces and determine the thickness. The distinct waveform of the bonding interface echo compared with the Cu/dielectric interface echo suggests that its acoustic response may also reflect the interfacial bonding conditions. The extracted thermal conductivity ranges from 1.14 to 1.75 $W \cdot m^{-1} \cdot K^{-1}$ for the bottom pad layer and from 1.76 to 4.34 $W \cdot m^{-1} \cdot K^{-1}$ for the top pad layer. The difference in thermal conductivity is consistent with the different Cu area fractions of the two layers. Furthermore, the thermal conductivity of the top pad layer increases significantly with the remaining thickness, possibly due to the thermal boundary resistance at the bonding interface, low thermal conductivity of SiCN interlayer, and increased electron scattering in thin Cu structures. Our work demonstrates a practical approach to investigate the thermal properties of buried hybrid bonding layers, providing experimental data for thermal design and modeling of 3D-ICs.

## 1. Introduction

The rapid growth of high-performance computing (HPC) and artificial intelligence (AI) has driven increasing demand for high integration density and improved system performance. As transistor scaling approaches its physical limits, three-dimensional integrated circuits (3D-ICs) have emerged as a promising approach for sustaining performance improvement [1]. Among the key technologies for 3D-ICs, hybrid bonding enables finer interconnect pitches than conventional micro-bumps, with reported pitch scaling below 10 μm and towards submicrometer scale [2][3][4]. Therefore, hybrid bonding has been widely adopted in various applications, including CMOS image sensors, high bandwidth memory (HBM), 3D NAND, and 3D V-cache [5][6][7].

Hybrid bonding has been reported to improve power delivery and signal routing by replacing micro-bumps with densely aligned Cu/dielectric interconnects, which provide fine-pitch connections and enable inter-tier metal sharing [8][9][10]. Beyond electrical advantages, the thermal properties of hybrid bonding layers are also critical as the power density of 3D-ICs continues to increase. Hybrid bonding is expected to provide low thermal resistance by replacing solder and underfill with a thin Cu/dielectric bonding layer [11][12][13]. Bulk copper has an intrinsic thermal conductivity of approximately 400 $W \cdot m^{-1} \cdot K^{-1}$, which is much higher than that of typical solder or polymer underfill materials [14][15].

Several simulation studies have evaluated the effective thermal conductivity of hybrid bonding layers based on Cu/dielectric finite element model and incorporated it into package-level simulations [16][17]. However, as the Cu feature size continues to shrink, size effects and thermal boundary resistance can increasingly affect the thermal properties of hybrid bonding layers [18][19][20]. Experimental studies have also characterized thermal transport in hybrid bonding structures, including the effective thermal properties of bonded layers and the thermal conductivity of individual Cu pads [21][22][23][24]. However, the Cu pad dimensions can differ between the upper and lower bonding layers, potentially leading to variations in thermal transport across hybrid bonding structures. More detailed thermal characterization is therefore needed to understand the depth-dependent thermal properties and provide reliable inputs for thermal design and modeling of related devices and packaging.

In this study, we employed time-domain thermoreflectance (TDTR) for depth-resolved thermal characterization of hybrid bonding layers. Shallow-angle polishing of the sample enabled measurements at different remaining thicknesses, while picosecond acoustic echoes identified the bottom pad/dielectric and top/bottom pad interfaces. Separate layer and equivalent layer models were then used in TDTR fitting to evaluate the thermal conductivities of the individual pad layers and the equivalent hybrid bonding layer, respectively.

## 2. Sample Preparation

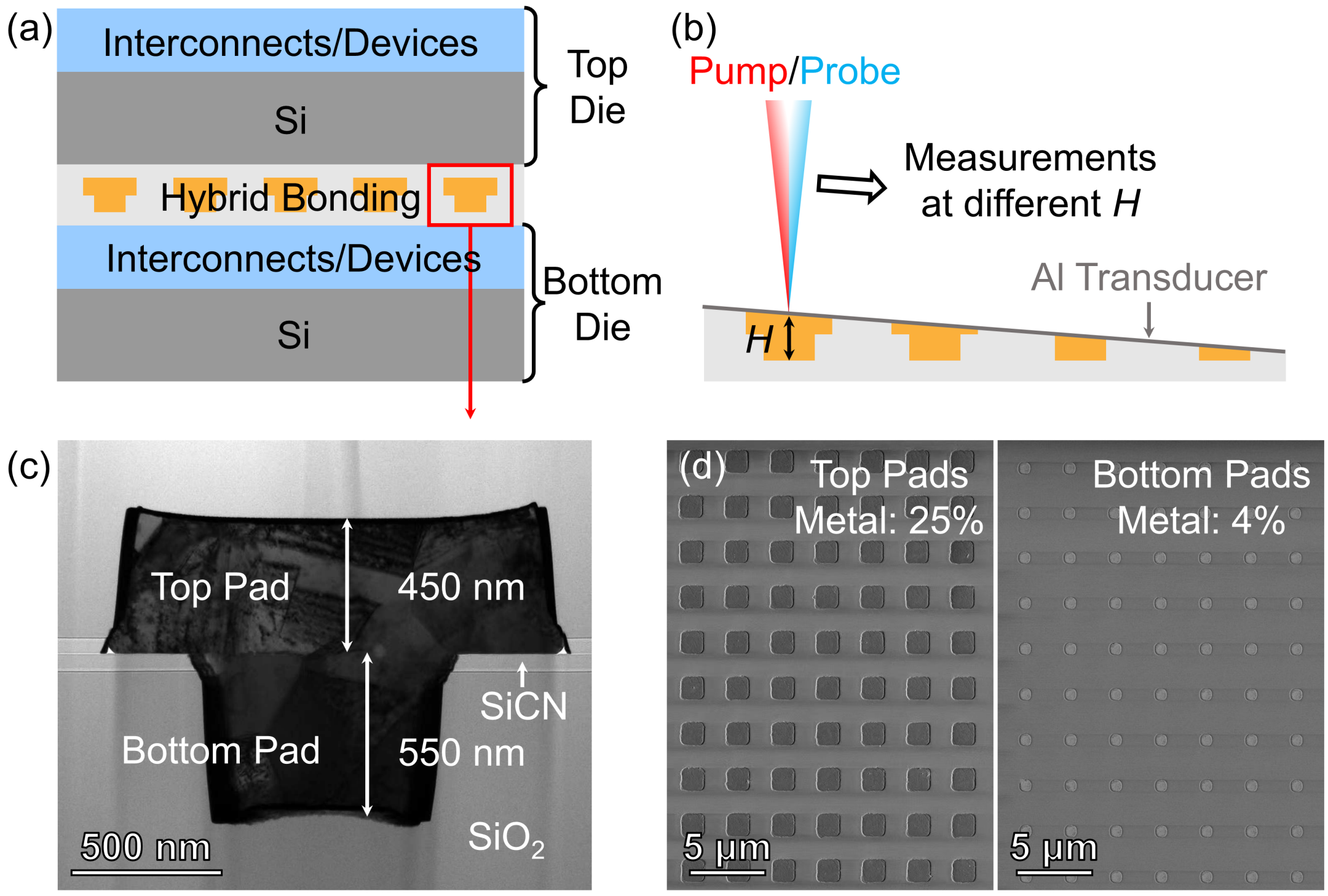

Fig. 1. Structure and preparation of the hybrid bonding sample. (a) Schematic cross section of the hybrid bonding structure within the stacked chip. (b) Polished sample geometry for TDTR measurements at different remaining thicknesses. (c) Cross-sectional TEM image of the hybrid bonding region. (d) Top-view SEM images of the top and bottom Cu pad arrays.

The chip sample consists of six dies connected through Cu/dielectric hybrid bonding layers in a face-to-back configuration. Figure 1(a) shows the local structure around the investigated bonding interface. The sample was polished at a shallow angle to enable TDTR measurements at different positions with varying remaining thicknesses, as illustrated in Fig. 1(b). An Al transducer layer was then deposited on the polished surface for TDTR measurements.

In TDTR, a modulated pump pulse train heats the Al transducer, where the absorbed optical energy is converted into heat. The resulting heat diffuses into the underlying layers, and the temperature response depends on the thermal properties of the multilayer structure. A time-delayed probe beam detects the temperature change from the reflectance variation, allowing the thermal response to be recorded as a function of pump-probe delay. The reflected probe signal is detected by a photodetector and analyzed using a lock-in amplifier to obtain the in-phase ($V_{in}$) and out-of-phase ($V_{out}$) components. The measured signals are then fitted using a multilayer thermal model to extract the effective thermal conductivity of the hybrid bonding layer [25][26].

The microstructure of the bonding region was examined by transmission electron microscopy (TEM), as shown in Fig. 1(c). The top and bottom Cu pads were embedded in the dielectric matrix, with thicknesses of approximately 450 and 550 nm, respectively. The dielectric region mainly consists of $SiO_2$, with a thin SiCN layer near the bonding interface. SiCN offers higher bonding energy and better Cu diffusion barrier than $SiO_2$ [27][28], but exhibits a lower thermal conductivity (0.6-0.8 $W \cdot m^{-1} \cdot K^{-1}$ [29]). Based on the scanning electron microscopy (SEM) images in Fig. 1(d), the Cu area fractions of the top and bottom pad arrays were estimated to be approximately 25% and 4%, using Otsu's thresholding method. The effective volumetric heat capacities of the pad layers were calculated as the volume-weighted average:

$$C_{eff}=f_{Cu}C_{Cu}+(1-f_{Cu})C_{dielectric} \qquad (1)$$

where $f_{Cu}$ is the Cu area fraction, and $C_{Cu}$ and $C_{dielectric}$ are 3.45×10$^6$ J·m$^{-3}$·K$^{-1}$ and 1.66×10$^6$ J·m$^{-3}$·K$^{-1}$, respectively. The volumetric heat capacity of $SiO_2$ was used to approximate that of dielectric, yielding $C_{top}$=2.11×10$^6$ J·m$^{-3}$·K$^{-1}$ and $C_{bottom}$=1.73×10$^6$ J·m$^{-3}$·K$^{-1}$.

## 3. Results and Discussion

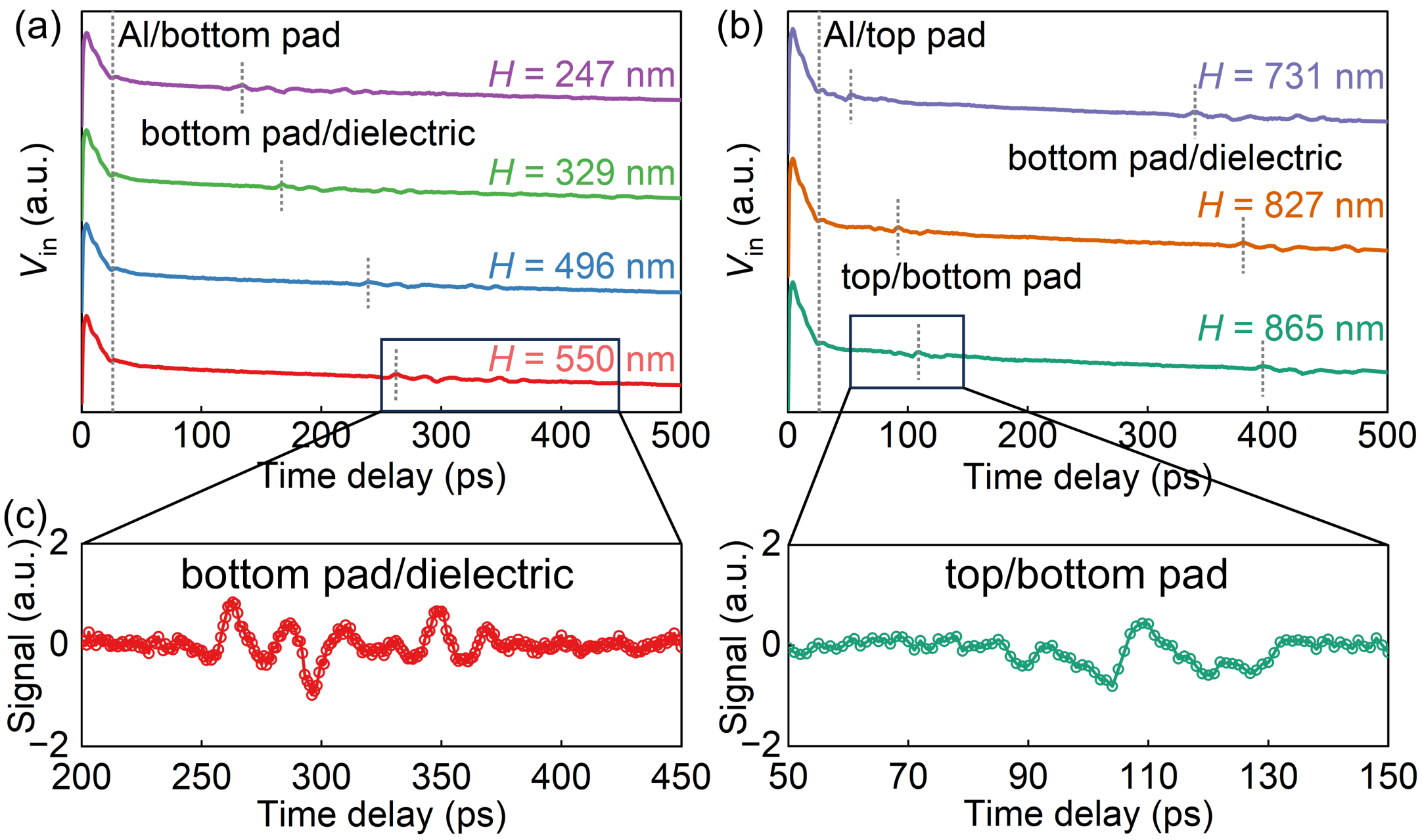

Fig. 2. Picosecond acoustic echoes measured at different remaining thicknesses $H$. (a) TDTR in-phase signals in the bottom pad region, showing the echoes of the Al/bottom pad and bottom pad/dielectric interfaces. (b) TDTR in-phase signals in the top pad region, showing the echoes of the Al/top pad, top/bottom pad, and bottom pad/dielectric interfaces. (c) Enlarged views of the acoustic echo signals after subtraction of the thermal background.

Figure 2 presents the picosecond acoustic responses measured at different remaining thicknesses *H* in the bottom and top pad regions. Absorption of the pump laser pulse in the Al transducer generates a longitudinal strain pulse through thermal expansion of lattice. As the strain pulse propagates through the underlying structure, it is reflected at the Cu/dielectric interface due to the difference in acoustical impedance [30]. The reflected pulses return to the surface and change the optical reflectivity, giving rise to acoustic echoes in the TDTR in-phase signals.

In the bottom pad region (Fig. 2(a)), the echoes are attributed to the Al/bottom pad and bottom pad/dielectric interfaces. As *H* increases from 247 to 550 nm, the echo shifts towards longer delay times, consistent with the increased acoustic propagation distance. In the top pad region (Fig. 2(b)), an additional echo reveals the bonding interface within the hybrid bonding structure. For clarity, the enlarged views in Fig. 2(c) show the acoustic echo signals of two interfaces after subtraction of the thermal background. Notably, the waveform of the top/bottom pad echo differs from that of the bottom pad/dielectric echo. As the latter mainly originates from the acoustical impedance mismatch between Cu and dielectric, the distinct waveform of the bonding interface echo suggests that its acoustic response may involve not only Cu/dielectric impedance mismatch but also the interfacial bonding conditions. Previous studies have reported acoustic echoes from the $SiO_x$-$SiO_x$ bonding interface [31][32]. Hao et al. [33] also found that interfacial roughness and imperfect bonding can modify the amplitude and shape of the echoes.

Using the picosecond ultrasonic technique, the remaining hybrid bonding thickness was estimated from the time interval between the echoes at the Al/pad and bottom pad/dielectric interfaces. The thickness was calculated as $H = v_{Cu}\Delta t/2$ , where $v_{Cu}$=4700 m/s is the longitudinal sound velocity of Cu [34], and $\Delta t$ is the time interval between two echoes. However, given the various possible reflection sources at the bonding interface discussed above, the bottom pad thickness used for top pad fitting was determined from TEM measurements rather than the acoustic echo. Fixing the layer thicknesses based on acoustic echoes and TEM measurements reduces parameter correlation in TDTR fitting, improving the reliability of thermal conductivity extraction.

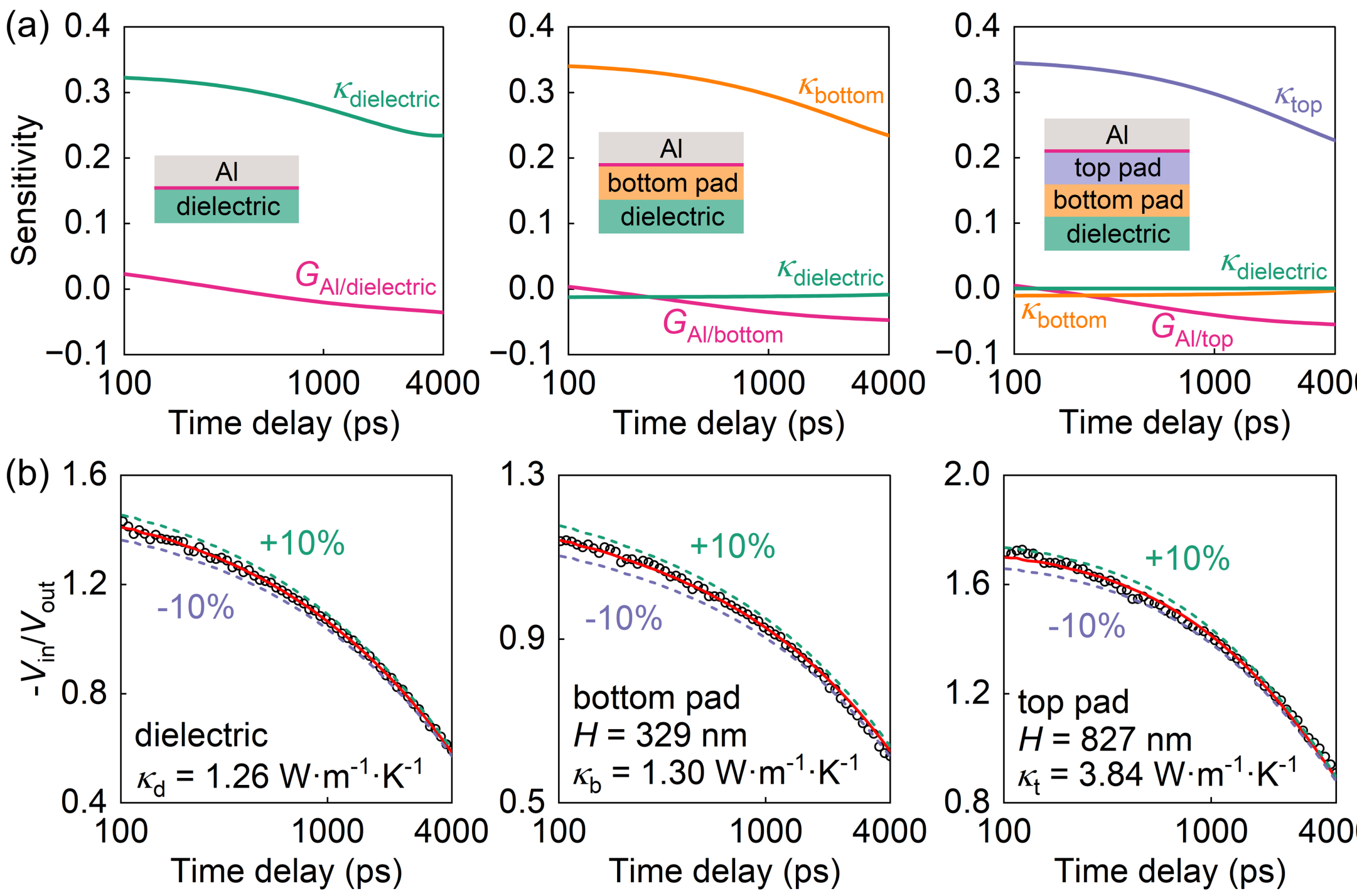

Fig. 3. TDTR sensitivity analysis and representative fitting results. (a) Sensitivities of the TDTR ratio ($-V_{in}/V_{out}$) to thermal conductivity $\kappa$ and thermal boundary conductance $G$ for the dielectric, bottom pad, and top pad regions. Insets illustrate the corresponding

thermal models. (b) Experimental TDTR ratios (symbols) and best fits (red solid lines) for the dielectric, bottom pad, and top pad regions. The pump and probe spot radii were 12 μm and 8 μm, respectively, allowing the measurements to capture the combined thermal response of multiple pads and the surrounding dielectric. The pump modulation frequencies were 10.08 MHz for the dielectric region and 5.49 MHz for the pad regions. Dashed lines show the calculated TDTR ratios with ±10% variations in thermal conductivity.

Figure 3(a) shows that the TDTR ratio is highly sensitive to the thermal conductivity of the layer directly beneath the Al transducer, whereas the sensitivities to the underlying layers are considerably lower. In particular, the sensitivities to the bottom pad and dielectric thermal conductivities are nearly zero in the top pad region, indicating their limited influence on the TDTR signals.

Measurements were performed sequentially in the dielectric, bottom pad, and top pad regions. In each region, the thermal conductivity of the layer directly beneath the Al transducer and its thermal boundary conductance with Al were fitted simultaneously, while the underlying thermal properties were fixed based on previous fits. The dielectric thermal conductivity was first determined to be 1.27 $W{\cdot}m^{-1}{\cdot}K^{-1}$ from the measurements at two randomly selected positions. For the top pad measurements, the bottom pad thermal conductivity was fixed at the average obtained at $H$ = 442, 496, and 550 nm. As shown in Fig. 3(b), the representative fitting results are 1.30 $W{\cdot}m^{-1}{\cdot}K^{-1}$ for the bottom pad at $H$ = 329 nm and 3.84 W·m$^{-1}$·K$^{-1}$ for the top pad at $H$ = 827 nm. Varying the fitted thermal conductivity by ±10% produces noticeable deviation from the best fits, confirming the sensitivity to the fitted thermal conductivity.

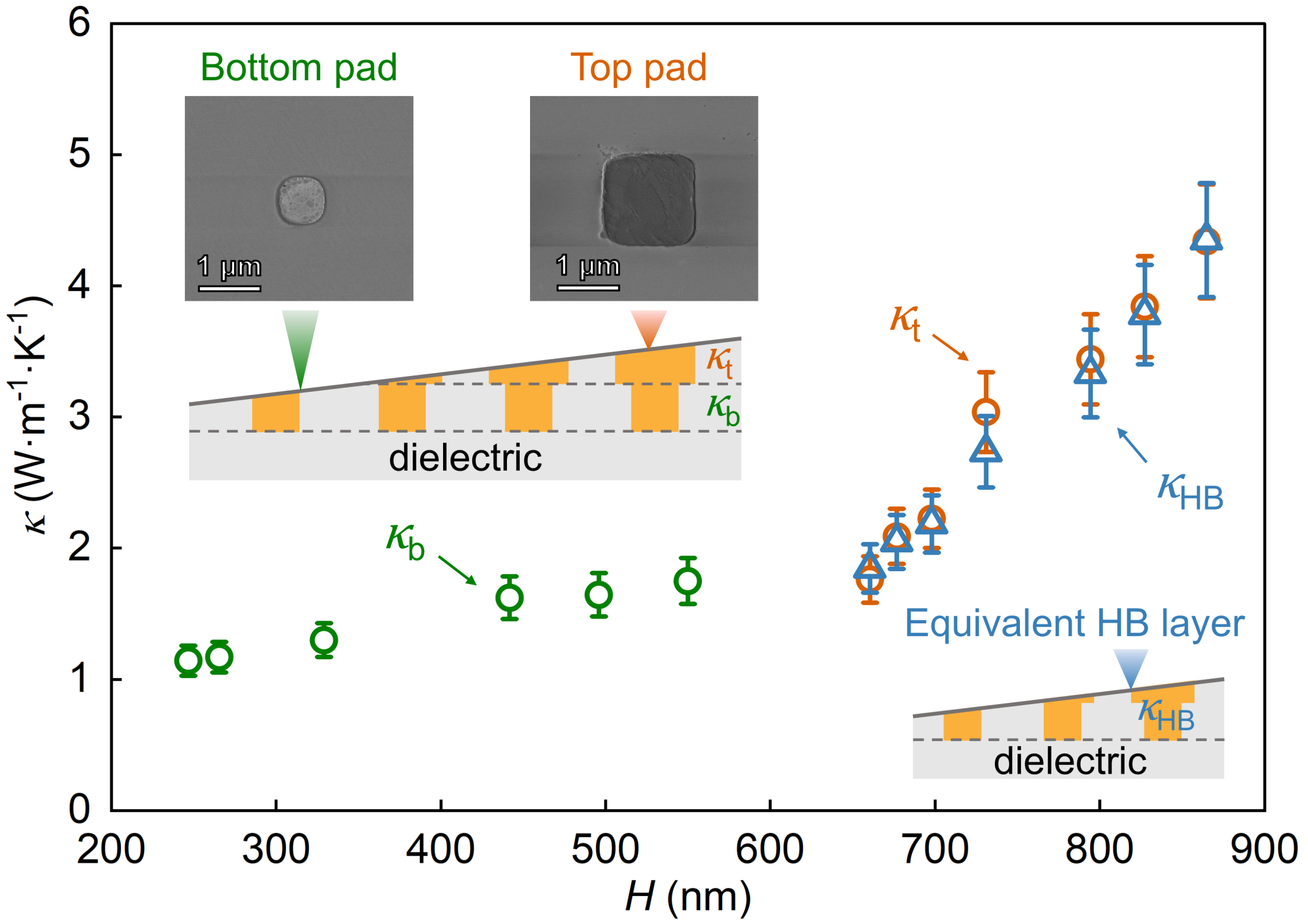

Fig. 4. Thermal conductivity as a function of the remaining thickness $H$. In the top pad region, orange circles represent the thermal conductivity of the top pad layer fitted by modeling the two pad layers separately. Blue triangles represent the effective thermal conductivity of the hybrid bonding (HB) layer. Insets show the SEM images of Cu pads and schematics of the thermal models. Error bars show a 10% systematic uncertainty of TDTR measurements.

Figure 4 summarizes the thermal conductivities at different remaining thicknesses extracted from two different fitting models. The thermal conductivity ranges from 1.14 to 1.75 $W \cdot m^{-1} \cdot K^{-1}$ for bottom pad layer and from 1.76 to 4.34 $W \cdot m^{-1} \cdot K^{-1}$ for top pad layer. The higher thermal conductivity of top pad layer is consistent with the larger Cu area fraction (25% versus 4%), as Cu has a much higher intrinsic thermal conductivity than the surrounding dielectric.

However, the measured values are much lower than those expected from the Cu area fractions alone. Assuming parallel heat conduction through Cu (~400 $W \cdot m^{-1} \cdot K^{-1}$) and dielectric (~1 $W \cdot m^{-1} \cdot K^{-1}$), the effective thermal conductivities would be approximately 100 $W \cdot m^{-1} \cdot K^{-1}$ for top pad layer and 20 $W \cdot m^{-1} \cdot K^{-1}$ for bottom pad layer. This discrepancy suggests that additional factors significantly affect heat transport in the bonding layers.

One possible factor contributing to the remarkably low thermal conductivity of the top pad region is the thermal boundary resistance (TBR) at the bonding interface. For a representative thermal boundary conductance (TBC) of 100 $MW \cdot m^{-2} \cdot K^{-1}$ [35], the corresponding resistance is approximately 10 $m^2 \cdot K/GW$. In comparison, the thermal resistance of the 450-nm-thick top pad layer (~3 $W \cdot m^{-1} \cdot K^{-1}$) is about 150 $m^2 \cdot K/GW$. Thus, the TBR alone is unlikely to account for the entire reduction in thermal conductivity. In addition, electron scattering at grain boundaries in microstructures may further reduce the thermal conductivity of Cu.

Notably, the thermal conductivity of the top pad layer increases significantly with the remaining thickness, which may arise from the thermal boundary resistance (TBR) of the bonding interface and the low thermal conductivity of the SiCN interlayer. The relative influence of TBR and SiCN is greater at smaller remaining thickness, leading to lower thermal conductivity. In addition, increased electron scattering in thinner Cu structures may reduce thermal conductivity [18], contributing to the thickness dependence. Furthermore, the fitted thermal conductivity of the top pad layer is similar to the equivalent thermal conductivity of the hybrid bonding layer. This is consistent with the low sensitivity to the underlying bottom pad shown in Fig. 3(a), indicating that the measured thermal response is dominated by the top pad layer.

## Conclusion

In this work, we achieved depth-resolved thermal characterization of hybrid bonding layers using TDTR combined with picosecond acoustic measurements. Distinct acoustic echoes were identified at the bottom pad/dielectric and bonding interfaces, with the latter potentially influenced by interfacial bonding conditions in addition to acoustical impedance mismatch. The results show that the top pad layer has a higher thermal conductivity than the bottom pad layer, consistent with its larger Cu area fraction. The thermal conductivity of the top pad layer also exhibits a significant thickness dependence, which may be related to bonding interface resistance, low conductivity of SiCN interlayer, and increased electron scattering in thin Cu structures. These findings provide experimental insights into thermal transport in HB structures and support thermal modeling of 3D-ICs. Future improvements in heat dissipation in hybrid bonding may focus on optimizing the Cu fraction, improving the Cu microstructure to reduce electron scattering, and minimizing TBR through improved bonding quality.

## Acknowledge

The authors acknowledge the financial support from the National Key Research and Development Program of China (Grant No.2024YFA1207901) and the National Natural Science Foundation of China (NSFC) (Grant Nos. 62574007, T2550270).